\documentclass[12pt]{article}

\usepackage{newtxtext,newtxmath}
\usepackage{float}

\usepackage{graphicx}

\usepackage[letterpaper,margin=1in]{geometry}

\renewenvironment{abstract}
	{\quotation}
	{\endquotation}

\date{}

\makeatletter
\renewcommand{\fnum@figure}{\textbf{Figure \thefigure}}
\renewcommand{\fnum@table}{\textbf{Table \thetable}}
\makeatother

\usepackage{scicite}

\usepackage{url}

\def\scititle{
	Magnon-mediated exciton-exciton interaction in a van der Waals antiferromagnet
}
\title{\bfseries \boldmath \scititle}

\author{
	Biswajit~Datta$^{1\ast\dagger}$,
	Pratap~Chandra~Adak$^{1\ast\dagger}$,
    Sichao~Yu$^{1,2\dagger}$,\and
	Agneya~V.~Dharmapalan$^{1,2,3}$,
    Siedah~J.~Hall$^{2,3}$,
    Anton~Vakulenko$^{4}$,\and
    Filipp~Komissarenko$^{5}$,
    Egor~Kurganov$^{4}$, 
    Jiamin~Quan$^{2}$,
    Wei~Wang$^{2}$, \and
    Kseniia~Mosina$^{6}$, 
    Zdeněk~Sofer$^{6}$, 
    Dimitar~Pashov$^{7}$, 
    Mark~van~Schilfgaarde$^{8}$,\and
    Swagata~Acharya$^{8}$, 
    Akashdeep~Kamra$^{9,10}$, 
    Matthew~Y.~Sfeir$^{2,3}$,\and
    Andrea~Alù$^{2,3,5}$, 
    Alexander~B.~Khanikaev$^{2,3,4}$, 
    Vinod~M.~Menon$^{1,2\ast}$\and
	\small$^{1}$Department of Physics, City College of New York, New York, NY 10031, USA.\and
	\small$^{2}$Department of Physics, Graduate Center of the City University of New York (CUNY),
    \small New York, NY 10016, USA.\and
    \small$^{3}$Photonics Initiative, CUNY Advanced Science Research Center, New York, NY 10031, USA.\and
    \small$^{4}$CREOL, The College of Optics and Photonics, University of Central Florida, Orlando, Florida 32816, USA,\and
    \small$^{5}$Department of Electrical Engineering, The City College of New York, New York, NY 10031, USA.\and
    \small$^{6}$Department of Inorganic Chemistry, University of Chemistry and Technology Prague, Prague, Czech Republic.\and
    \small$^{7}$Theory and Simulation of Condensed Matter, King's College London, The Strand, London WC2R2LS, UK.\and
    \small$^{8}$Materials Science Center, National Renewable Energy Laboratory, Golden, Colorado 80401, USA.\and
    \small$^{9}$Department of Physics, Rheinland-Pfälzische Technische Universität (RPTU),\and 
    \small Kaiserslautern-Landau, Kaiserslautern, Germany.\and
    \small$^{10}$Departamento de Física Teórica de la Materia Condensada and Condensed Matter Physics Center (IFIMAC),\and
    \small Universidad Autónoma de Madrid, E- 28049 Madrid, Spain.\and
	\small$^\ast$Corresponding authors. Email: bdatta@ccny.cuny.edu, padak@ccny.cuny.edu, vmenon@ccny.cuny.edu\and
	\small$^\dagger$These authors contributed equally to this work.
}

\begin{document} 

\maketitle

\begin{abstract} 
Excitons are fundamental excitations that govern the optical properties of semiconductors. Interacting excitons can lead to various emergent phases of matter and large nonlinear optical responses. In most semiconductors, excitons interact via exchange interaction or phase space filling. Correlated materials that host excitons coupled to other degrees of freedom offer hitherto unexplored pathways for controlling these interactions. 
Here, we demonstrate magnon-mediated excitonic interactions in CrSBr, an antiferromagnetic semiconductor. 
This interaction manifests as the dependence of exciton energy on exciton density via a magnonic adjustment of the spin canting angle. 
Our study demonstrates the emergence of quasiparticle-mediated interactions in correlated quantum materials, leading to large nonlinear optical responses and potential device concepts such as magnon-mediated quantum transducers.
\end{abstract}

\noindent
Interactions mediated by quasiparticles are often responsible for novel phenomena, such as phonon-mediated electronic interactions in superconductors or polaronic effects in solar cells.
In semiconductors, exciton-exciton interactions significantly influence optical properties and nonlinearity required for advanced photonics and quantum information processing. 
Furthermore, exciton-exciton interactions lead to quantum many-body physics~\cite{reganEmergingExcitonPhysics2022b} and emergent phases such as Bose-Einstein condensation\cite{snokeSpontaneousBoseCoherence2002, wangEvidenceHightemperatureExciton2019b, dengCondensationSemiconductorMicrocavity2002, kasprzakBoseEinsteinCondensation2006a}, superfluids\cite{amoSuperfluidityPolaritonsSemiconductor2009, liExcitonicSuperfluidPhase2017a}, excitonic insulators\cite{jeromeExcitonicInsulator1967, cercellierEvidenceExcitonicInsulator2007,kogarSignaturesExcitonCondensation2017}, and Wigner crystals\cite{joglekarWignerSupersolidExcitons2006}.
Magnetic semiconductors can simultaneously host excitons and magnons (quantized magnetic excitations), which typically remain independent due to their disparate energy scales.
Here, we demonstrate a new mechanism where magnons mediate attractive nonlinear interactions between excitons~\cite{johansen_magnon-mediated_2019}.
In most known excitonic materials, excitons interact through mechanisms such as phase-space filling or short-range exchange interactions\cite{axtNonlinearOpticsSemiconductor1998}. In certain cases, long-range dipolar interactions dominate, as seen with dipolar excitons\cite{liDipolarInteractionsLocalized2020}, or the long-range Coulomb interactions play a key role, as with charged excitons (trions). 
In all of these scenarios, the excitons directly interact with each other. 
In contrast, the magnon-mediated excitonic interaction reported here offers an indirect pathway via magnon-induced changes in the electronic structure. 

We use CrSBr, a magnetic van der Waals (vdW) semiconductor with unique magneto-electric coupling~\cite{wilson_interlayer_2021,bae_exciton-coupled_2022,diederich_tunable_2023,dirnberger_magneto-optics_2023, brennan_important_2024}, to demonstrate this magnon-mediated excitonic interaction.
CrSBr possesses an A-type antiferromagnetic (AFM) structure with a Néel temperature of 132~K (bulk) to $\sim$150~K (few-layers)~\cite{telford_layered_2020, wang_electrically-tunable_2020, lee_magnetic_2021, scheie_spin_2022}. 
Unlike transition metal dichalcogenides, CrSBr maintains a direct bandgap~\cite{PhysRevB.107.235107,watson2024giant} and exhibits excitonic photoluminescence (PL) across all thicknesses~\cite{wilson_interlayer_2021, shao2024exciton}. 
Furthermore, CrSBr supports strongly bound excitons with large oscillator strength, facilitating self-hybridized polaritons in thick flakes aided by the large refractive index~\cite{dirnberger_magneto-optics_2023, wang_magnetically-dressed_2023}.
A distinct feature of CrSBr is its magnetic field-tunable exciton energy, governed by the canting angle between the magnetizations of its two antiferromagnetically coupled sublattices~\cite{wilson_interlayer_2021}.
Notably, the temporal modulation of this canting angle by magnons induces oscillations in exciton absorption, providing direct optical access to magnon dynamics using time-resolved optical spectroscopy~\cite{bae_exciton-coupled_2022,diederich_tunable_2023, sun_dipolar_2024}. 

In this study, we employ pump-power-dependent exciton absorption spectroscopy to probe magnon-mediated exciton-exciton interactions in CrSBr. This interaction is evidenced by a redshift in the exciton energy with increasing exciton density.
CrSBr has two primary excitons ($A$ and $B$) at 1.36~eV and 1.77~eV, respectively. 
Our experiments reveal a large redshift in the $B$ exciton energy with increasing fluence, whereas the $A$ exciton shows negligible nonlinearity. 
Notably, the redshift is maximized at an intermediate magnetic field ($0<B<B_\text{sat}$) that disrupts the spin ordering, and is minimized at the extremes (AFM at $B=0$\,T and ferromagnetic (FM) at $B >B_\text{sat}$).
This observation suggests that a finite canting angle is necessary for the shift, underscoring the crucial role of exciton-magnon coupling, which strengthens with the noncollinearity between the two magnetizations, as confirmed by pump-probe measurements.
The differing nonlinear responses of the two excitons stem from their contrasting magneto-optical behavior due to distinct characters of the $A$ and $B$ excitons, as supported by first-principles many-body perturbation theory.
Our model of magnon-mediated excitonic interaction accurately captures the non-monotonic dependence of the excitonic nonlinearity on the magnetic field.

\begin{figure}
    \centering
	\includegraphics[width=15cm]{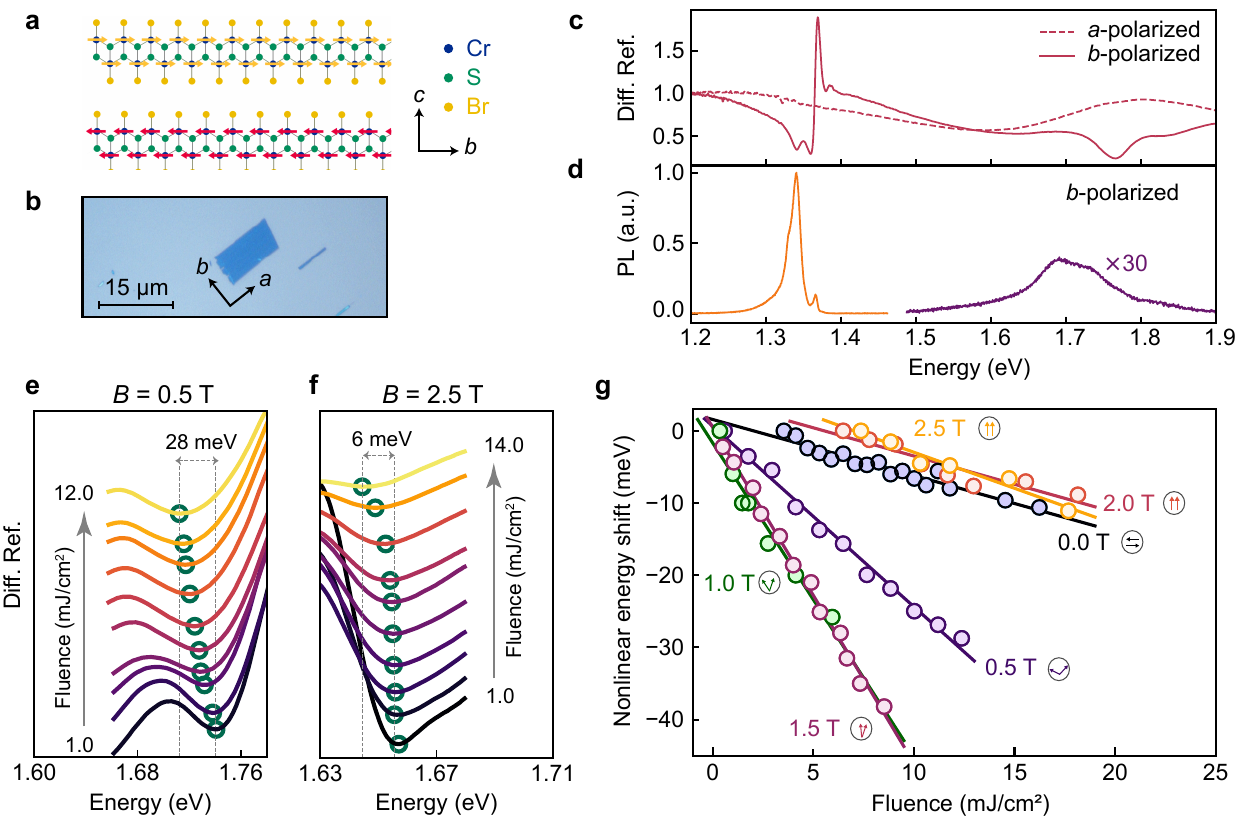}
        \caption{  \textbf{Excitons in CrSBr and nonlinear optical response.}
        (\textbf{a})~Crystal structure of CrSBr projected on the $bc$-plane showing CrS layers sandwiched between Br atoms. The arrows indicate spin alignment in the AFM state.
        (\textbf{b})~Microscopic image of rectangular-shaped CrSBr flakes exfoliated on a 90~nm SiO$_2$/Si substrate, oriented with the long side along $a$-axis due to strong anisotropy.
        (\textbf{c})~Differential reflectivity vs. energy plots for a 3~nm thick CrSBr sample exfoliated on a SiO$_2$/Si substrate under two different polarization states. 
        The spectrum under $b$-polarization reveals distinct dips at 1.36~eV and 1.77~eV corresponding to $A$ and $B$ excitons.
        (\textbf{d})~Photoluminescence (PL) measurements: PL near the $A$ exciton is from the same sample, while PL near the $B$ exciton is from a 7~nm flake on a distributed Bragg reflector (DBR) centered at 700~nm to enhance the PL intensity.
        (\textbf{e,f})~Differential reflectivity vs. energy plots around the $B$ exciton as a function of fluence using pulsed laser polarized along the $b$-axis under various magnetic fields perpendicular to the flake. Plots in \textbf{e} and \textbf{f} are for $B=0.5$~T and 2.5~T, respectively. 
        (\textbf{g})~Nonlinear energy shift vs. fluence plots for the $B$ exciton, derived from resonant differential reflectivity measurements at different fields. The energy shifts are plotted relative to its low-fluence value at each $B$.
        }
        \label{fig:fig1}
\end{figure}

CrSBr possesses an orthorhombic crystal structure with layers stacked in the $c$-direction, held together by vdW forces~\cite{guo_chromium_2018, telford_layered_2020, wilson_interlayer_2021}.
Each layer consists of two staggered rectangular planes of CrS sandwiched between Br atoms on both sides (Fig.~\ref{fig:fig1}a).
Below the Néel temperature, the material exhibits A-type AFM ordering, where the magnetic moments of successive layers are oriented in opposite directions along the $b$-axis.
The significant crystallographic anisotropy of CrSBr results in typically rectangular-shaped exfoliated flakes, with the longer (shorter) side along the crystal axis $a$ ($b$) (Fig.~\ref{fig:fig1}b).

Fig.~\ref{fig:fig1}c shows the differential reflectivity of a 3 nm thick CrSBr flake measured at 5~K under zero magnetic field, using incoherent white light from a halogen source~\cite{methods}. 
For light polarized along the $b$-axis, the reflectivity spectrum shows distinct absorption features at 1.36\,eV, attributed to the tightly bound $A$ exciton. 
The broader dip observed at 1.77~eV corresponds to the $B$ exciton. These features disappear for light polarized along the $a$-axis, indicating that the transition dipole moments of both excitons are predominantly oriented along the $b$-axis.
This anisotropy is further evidenced in the photoluminescence (PL) spectra shown in Fig.~\ref{fig:fig1}d, which is strongly polarized along the $b$-axis, consistent with previous reports. 
Additionally, we detect PL from the $B$ exciton, enhanced by employing a distributed Bragg reflector (DBR) centered at 700 nm as the substrate. 
However, the PL intensity of the $B$ exciton is much weaker than that of the $A$ exciton.
The differential reflectivity in some flakes features an additional dip near the $B$ exciton energy, possibly due to two nearly degenerate excitons with energy separation dependent on the flake thickness. Forward and backward magnetic field sweeps exhibit a pronounced hysteresis in differential reflection near both the A and B exciton resonances, as expected due to the underlying coupled magnetic order, see Fig. S5.


A key indicator of exciton-exciton interactions is the nonlinear optical response. 
Repulsive interactions cause an energy blueshift, whereas attractive interactions result in a redshift with increasing exciton density.
To study exciton-exciton interactions in CrSBr, we perform pump-power-dependent resonant reflectivity experiments, probing excitonic absorption as a function of fluence at different magnetic fields applied along the $c$-axis.
We use a supercontinuum pulsed laser with a frequency window limited around the exciton's absorption dip using a bandpass filter in the input path (see Fig. S1)~\cite{methods}.
Figures~\ref{fig:fig1}E and \ref{fig:fig1}F display differential reflectivity measurements around the $B$ exciton as a function of fluence for two different magnetic fields.
The magnetic field induces a redshift in exciton energy that is evidenced even at low fluence.
Intriguingly, we observe an additional large redshift that increases with fluence. 
Fig.~\ref{fig:fig1}g summarizes the results, showing the energy shift relative to its low-fluence value for different magnetic fields.
The results indicate that the exciton energy shifts linearly with fluence beyond a certain threshold. 
Interestingly, the redshift varies non-monotonically with the magnetic field.
The redshift is small at zero field and $B>B_{c,\text{sat}}=2.0$~T, when the spins are frozen along one axis, whereas it is maximal at a finite magnetic field $0<B<B_{c,\text{sat}}$.

In contrast, the $A$ exciton exhibits very small fluence-induced shifts at all magnetic fields (Fig. S4a), suggesting weaker exciton-exciton interactions than the $B$ exciton.  
Additionally, experiments with varying laser repetition rates confirm that the observed redshifts are independent of repetition rate, ruling out thermal effects (Fig. S4b). 

The observation of magnetic field-dependent nonlinear responses underscores a direct link between exciton-exciton interactions in CrSBr and the magnetic order, challenging a potential role of Coulomb or exchange interactions alone. 
Exchange interactions between excitons lead to a blueshift, while Coulomb and dipolar interactions\cite{datta2022highly, louca2023interspecies} can result in both blueshift and redshift of the exciton energy.
However, neither mechanism can account for the unique magnetic field dependency of the redshift observed in CrSBr.

\begin{figure}
    \centering
	\includegraphics[width=15cm]{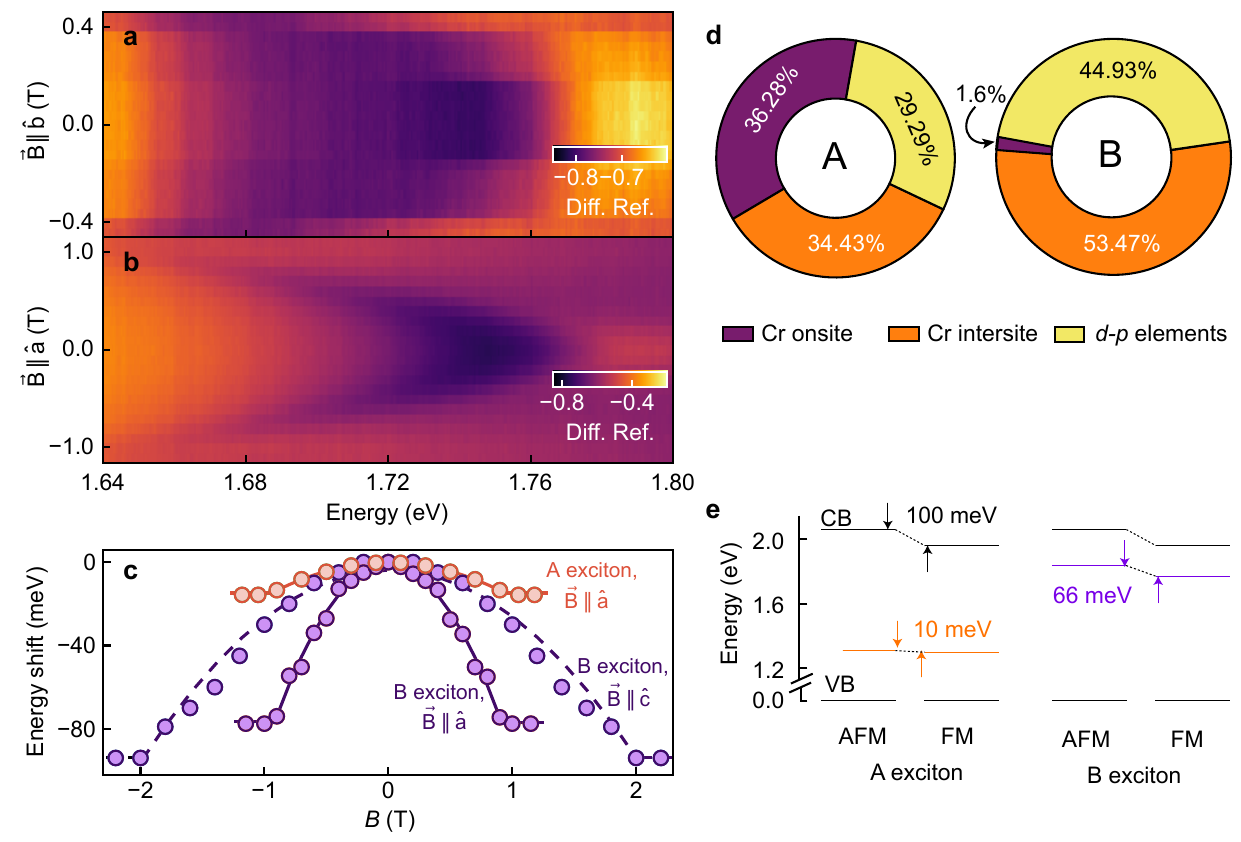}
        \caption{\textbf{Magnetic field-induced shifts in exciton energy.}
        (\textbf{a,b})~Color-scale plots of the differential reflectivity for the $B$ exciton as a function of energy and magnetic field applied along $b$- and $a$-axes, respectively. For $B\parallel b$, the exciton energy exhibits abrupt redshifts due to spin-flip transitions under the magnetic field.
        For $B\parallel a$, exciton energy shifts gradually due to progressive spin canting.
        (\textbf{c})~Plots showing the extracted energy shifts induced by the magnetic field. The maximum energy shift for the $B$ exciton, when the magnetic field is applied along the $a$ axis, is approximately $80$~meV—about five times larger than that observed for the $A$ exciton.
        (\textbf{d})~Wavefunction of the $A$ and $B$ excitons are decomposed in the atomic basis. In purple is shown the Frenkel component, i.e., the probability of having both the electron and the hole (forming the excitons) on the same Cr atom, while the orange represents the probability of having electron and hole on different Cr atoms. While the $A$ exciton has a rather large onsite Frenkel component, the Frenkel component nearly vanishes for the $B$ exciton. 
        (\textbf{e})~Schematic summarizing the theoretical findings, explaining the differences in magnetic field-induced energy shifts between the two excitons.
        The transition from AFM to FM order changes the bandgap by 100 meV. 
        Correspondingly, the binding energy of $B$ exciton changes by 34 meV, resulting in a 66 meV change in the exciton energy.
        The $A$ exciton energy shows a smaller change of  10 meV.
        }
        \label{fig:fig2}
\end{figure}

To further explore the relation between exciton-exciton interactions and magnetic order, we study the magnetic field effects on excitons. 
Figures~\ref{fig:fig2}A and \ref{fig:fig2}B present color-coded plots of differential reflectivity around the $B$ exciton energy, with magnetic fields applied along the $b$ and $a$ axes (see Fig. S2 for $A$ exciton energy and Fig. S3 for measurements of $B\parallel c$).
We use broadband low-power incoherent white light polarized along the $b$-axis for these measurements~\cite{methods}. 
Notably, an abrupt shift in exciton energy occurs at the saturation field of $B_{b,\text{sat}}=0.4$~T, when the field is applied along the $b$-axis (easy axis). 
As the field increases along $a$ and $c$ axes, exciton energy shifts gradually until reaching the saturation magnetic fields of $B_{a,\text{sat}}=1.0$~T and $B_{c,\text{sat}}=2.0$~T, respectively.
Beyond these saturation fields, the exciton energy remains constant.

Fig.~\ref{fig:fig2}c shows the evolution of exciton energy for both $A$ and $B$ excitons with $B\parallel a$ and $B\parallel c$. 
For both excitons, the energy shifts adhere to the equation $E_X-E_{X,0} \propto B^2$, till $B=B_\text{sat}$, where $E_X$ is the exciton energy at a finite $B$ and $E_{X,0}$ is at $B=0$. 
At a finite perpendicular magnetic field, the magnetization vectors of neighboring layers cant away from their equilibrium positions and the canting angle $\theta$ between the two magnetization vectors deviates from $(\pi) 0$, which denotes the (anti)ferromagnetic configuration. 
The field dependence of exciton energy can be formulated as $E_X-E_{X,0}=-\Delta_B\cos^2(\theta/2)$, where $\Delta_B$ is the maximum redshift.
Although  magnetic field-induced energy shifts are observed for both excitons, the shift for the $B$ exciton is markedly larger-- decreasing by 80~meV from the AFM to the FM state, compared to a 15~meV decrease for the $A$ exciton.

To explain the larger energy shift for the $B$ exciton with the magnetic field compared to the $A$ exciton, we compute electronic eigenfunctions in the presence of excitonic correlations using an \textit{ab initio} self-consistent many-body perturbative approach, $\mathrm{QSG\hat{W}}$\cite{Cunningham2023} (Supplementary Note IV).
Fig.~\ref{fig:fig2}d shows the wavefunction decomposition across different atomic sites in CrSBr for the two excitons.
The $B$ exciton is largely delocalized (extending up to $\sim$4 nm, Fig. S9) and has predominantly Wannier character.
In contrast, the $A$ exciton is mostly localized near a single Cr site (extending up to $\sim$1 nm, Fig. S7) and displays predominantly Frenkel character, well described by ligand-field theory akin to the Cr$^{3+}$ multiplet lines in Ruby.

Fig.~\ref{fig:fig2}e illustrates the calculated bandgaps in the AFM and FM states, along with the calculated exciton energies (see Figures S6 and S8 for the band structures).
In the AFM state, hybridization between anti-aligned planes is spin-forbidden, creating an energy barrier for electrons at the conduction band minimum~\cite{wilson_interlayer_2021}. 
In the FM state, with aligned spins, the potential is uniform across planes, reducing the energy barrier and consequently the bandgap.  
Detailed calculations reveal an AFM bandgap of $\sim$2.1\,eV~\cite{watson2024giant}, and an FM bandgap $\sim$0.1\,eV smaller.
The $B$ exciton energy tracks the conduction band, as expected for Wannier excitons from the classical effective mass description generalized to an anisotropic case.
The 0.1\,eV bandgap reduction results in a 66~meV reduction in the $B$ exciton energy.
Note that this tracking is incomplete due to the $B$ exciton's partial delocalization and a slight reduction in binding energy in the FM state.
Conversely, the $A$ exciton, with its largely Frenkel-like character, exhibits a weaker dependence on the host band structure, as the notion of binding relative to a band edge state is less relevant in the ligand-field picture.
Thus, the $A$ exciton energy shows minimal change of $10$~meV due to the AFM to FM transition.

\begin{figure}
    \centering
	\includegraphics[width=6.4cm]{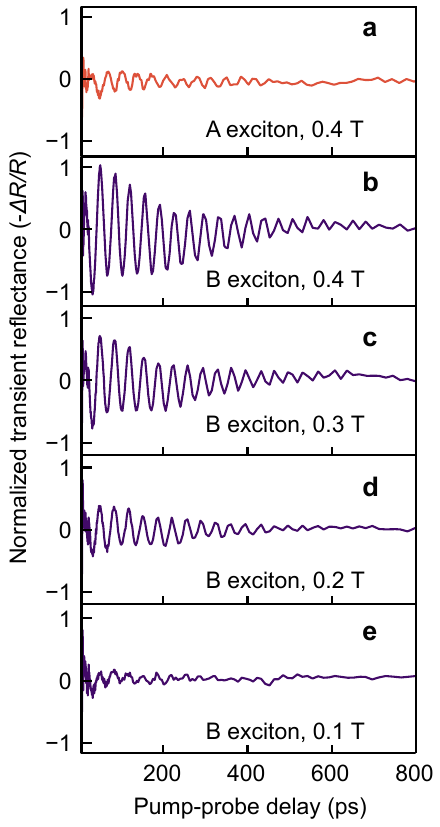}
        \caption{\textbf{Exciton-magnon coupling.}
        (\textbf{a,b})~Transient reflectance as a function of pump-probe delay measured on a 50~nm thick flake at a temperature of 6~K with a magnetic field of 0.4~T applied along the $a$-axis for probe energy corresponding to $A$ and $B$ excitons, respectively. 
        The oscillation amplitude is much higher for the $B$ exciton than the $A$ exciton, indicating stronger exciton-magnon coupling for the $B$ exciton.
        (\textbf{c-e})~Transient reflectance for the $B$ exciton for three additional magnetic field values. The oscillation amplitude changes with the magnetic field, indicating field-dependent exciton-magnon coupling. All the oscillations are normalized to the maximum magnitude of the oscillation of the B exciton at 0.4~T.
        }
        \label{fig:fig3}
        
\end{figure}

The dependence of exciton energy on spin alignment discussed above leads to exciton-magnon coupling, manifested as time-dependent oscillations in exciton energy at the magnon frequency$\sim$28 GHz~\cite{brennan_important_2024}.
Figures~\ref{fig:fig3}A and \ref{fig:fig3}B show such oscillations for $A$ and $B$ excitons, respectively, in a 50~nm thick flake at 0.4~T applied along the $a$-axis measured using transient reflectivity measurements~\cite{methods}.
The larger oscillation amplitude observed for the $B$ exciton, consistent with its greater magnetic field-induced energy shift, indicates stronger exciton-magnon coupling of the $B$ exciton compared to the $A$ exciton. 
This correlation between the strength of exciton-magnon coupling and the observed nonlinearity for the two excitons suggests that exciton-magnon coupling plays a crucial role in exciton-exciton interactions.
Figures~\ref{fig:fig3}B-E further demonstrate this by showing oscillations for the $B$ exciton at different magnetic field values. 
The oscillation amplitude increases with increasing magnetic field, establishing a clear field dependence of the exciton-magnon coupling. 
Previous studies~\cite{diederich_tunable_2023} have reported a similar field-dependence for the $A$ exciton, with maximum coupling occurring at an intermediate magnetic field corresponding to a canting angle of $\pi/2$.
The similar non-monotonic field dependence of the excitonic nonlinearity observed in Fig.~\ref{fig:fig1}g further supports the role of exciton-magnon coupling in mediating the excitonic interactions.


We now discuss how the exciton-magnon coupling manifests in the excitonic interaction.
Any deviation of the canting angle from $\pi$ increases the excitons' ability to disperse across different layers, reducing their kinetic energy.
Consequently, exciton energy is dependent on $\theta$.
Ordinarily, $\theta$ is adjusted to minimize magnetic energy alone.
However, when exciton energy is also dependent upon $\theta$, the optimal magnetic configuration, i.e., the canting angle that minimizes the total energy (combining both magnetic and exciton energies) varies with the number of excitons. 
As a result, the energy difference between states with varying numbers of excitons is not constant but decreases as the number of excitons increases, with corresponding adjustments in $\theta$.
This redshift is tantamount to an exciton-exciton attraction mediated by their coupling to the magnetic order and associated magnonic excitations.

\begin{figure}
    \centering
	\includegraphics[width=15cm]{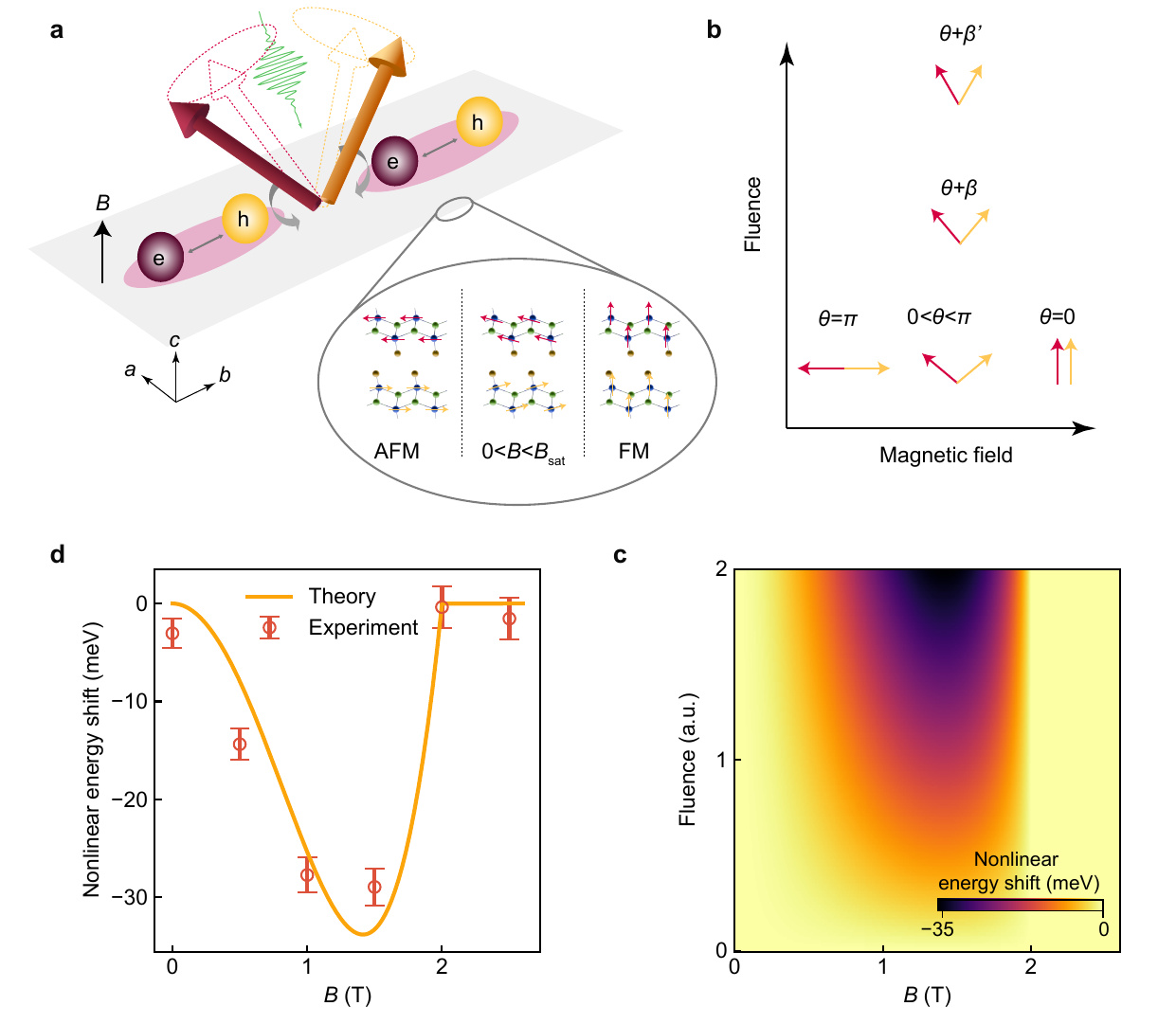}
        \caption{\textbf{Magnon-mediated exciton-exciton interactions.}
        (\textbf{a})~Schematic of exciton-exciton interactions in CrSBr mediated by magnons. 
        Under a finite magnetic field applied perpendicular to the flake, the magnetization vectors of two sublattices (i.e., neighboring layers) tilt with angle $\theta$ between them. 
        Due to the exciton-magnon coupling in noncollinear configurations, this angle depends on the exciton density. 
        The inset shows atomic arrangements in CrSBr crystals projected on the $bc$-plane, along with spin alignments at different magnetic fields, including AFM and FM orders.
        (\textbf{b})~Schematic depiction of the effect of magnetic field and fluence on the canting angle between two sublattice magnetization vectors.  Upon application of a perpendicular magnetic field, $\theta$ gradually changes from 0 (AFM) to $\pi$ (FM). 
        At a finite magnetic field, the magnon-induced change $\beta$ in $\theta$ changes with fluence, thereby affecting the exciton energy.
        (\textbf{c})~Color-scale plot of calculated nonlinear part of the exciton energy shifts as a function of fluence and magnetic field for the $B$ exciton.
        (\textbf{d})~A line slice from (c)  showing nonlinear energy shifts as a function of the magnetic field at a constant fluence.
        The experimental data points correspond to the energy shifts at fluence = 6 \text{mJ/cm$^2$} extracted from \ref{fig:fig1}g. 
        }
        \label{fig:fig4}
\end{figure}

We explain the magnetic field-dependent excitonic nonlinearity in CrSBr using a model of magnon-mediated exciton-exciton attraction, similar to phonon-mediated exciton interactions\cite{yazdaniCouplingOctahedralTilts2024}.
As depicted in Fig.~\ref{fig:fig4}a,  magnonic modes admit deviation of the sublattice magnetizations from their magnetic field-determined directions in the presence of a perpendicular magnetic field. 
This results in a small exciton-induced adjustment, $\beta$, in the canting angle $\theta$.
The total energy of the system, combining magnetic and excitonic contributions under an applied magnetic field $B$, is given by, $E_\text{total}(N) = E_\text{cons}(B) + \frac{1}{2} \chi \beta^2 +N E_\text{X} $.
Here, $E_\text{cons}(B)$ includes all other energy contributions not explicitly mentioned, and $\chi$ characterizes an energy corresponding to the considered magnon mode expressed via the $\beta^2$ term as a harmonic oscillator.
As the fluence of the pulsed laser increases, elevating the number of excitons ($N$), the system adjusts $\beta$ to minimize its total energy (Fig.~\ref{fig:fig4}b). 
Assuming magnetization responds sufficiently fast to changes in exciton population and $|\beta|\ll |\theta|$, we derive the minimum system energy, $E_\text{min}(N)$ (Supplementary Note V).
The energy of the absorbed photon when transitioning from a state with $N-1$ to $N$ excitons is then obtained by
\begin{equation}\label{eqn:interaction}
    \hbar \omega = E_\text{min}(N)-E_\text{min}(N-1)= E_{X,0} - \Delta_B \cos^2\left(\frac{\theta}{2}\right) - (2N-1)\frac{\Delta^2_B  \sin^2\theta}{8\chi}.
\end{equation}

Equation~\ref{eqn:interaction} succinctly captures the key findings of our experiments.   
The second term represents the overall exciton energy shift due to the magnetic field, observable in Figs.~\ref{fig:fig2}A-C. 
The third term quantifies the fluence-dependent shift in exciton energy, notably proportional to the square of $\Delta_B$. 
Given that $\Delta_B$ for the $A$ exciton is five times smaller than that for the $B$ exciton, this relationship explains the relatively minor fluence-induced shift observed for the $A$ exciton.
Moreover, for $N\gg 1$, the fluence-induced shift scales linearly with $N$, aligning with our experimental results shown in Fig.~\ref{fig:fig1}g.
Fig.\ref{fig:fig4}c illustrates the calculated nonlinear energy shift as a function of both magnetic field and fluence, employing Equation~\ref{eqn:interaction}.
The exciton number-dependent energy shift (third term) peaks at an intermediate magnetic field where $\theta = \pi/2$ and disappears at $\theta = 0, \pi$, correlating with minimal observed energy shifts in both FM and AFM states.
In Fig.\ref{fig:fig4}d, we further plot the energy shifts as a function of the magnetic field at a constant fluence along with experimental data from Fig.\ref{fig:fig1}g.
The trends predicted by our theoretical model of magnon-mediated exciton-exciton interactions are closely aligned with experimental observations.  

In materials such as layered perovskites, TMDCs, ZnO and GaN-based dilute magnetic semiconductors (DMSs), excitonic shifts under high magnetic fields are typically minor and driven by mechanisms like the Zeeman shift, diamagnetic effect, and magneto-Stark shift~\cite{chaves_tunable_2021, zipfel_spatial_2018, pacuski_excitonic_2006}.
In contrast, the large field-induced shift in CrSBr is unique, originating from field-dependent inter-layer hopping governed by the canting angle.
Since the canting angle also depends on magnonic modes, magnons and excitons are no longer independent.
With an increasing number of excitons, the system readjusts the spin alignment by influencing the magnon modes to minimize the energy. 
The resultant reduction in canting angle reduces the average exciton energy, thus mediating exciton-exciton interaction. 

This interaction exhibits two notable aspects beyond the large nonlinearity. 
Firstly, the unusually large shift in exciton energy indicates a long-range interaction as opposed to more conventional short-range exchange interactions, which require overlapping exciton wavefunctions. 
While dipole-dipole exciton interactions are long-range, they are prominently influenced by charge screening. 
We speculate that the magnon-mediated interaction revealed here will be independent of such screening. 
Although our simple model assuming a single homogeneous magnon mode captures the essence of the interaction, a detailed calculation involving different magnon modes and considering symmetries is necessary to quantify the range accurately. 
Another intriguing aspect is that, unlike direct exciton-exciton interactions, excitonic interactions mediated by another quasiparticle provide tunability of the excitonic nonlinearity using external knobs such as the magnetic field.

In summary, our study demonstrates that magnons in CrSBr not only affect exciton energy but also mediate exciton-exciton interactions. Supported by theoretical insights, our work advances the understanding of exciton interactions in magnetic vdW materials and suggests novel applications in realizing tunable nonlinear optics and quantum technologies.


\clearpage

\clearpage

\clearpage

\subsection*{Methods}

\subsubsection*{Fabrication details}
CrSBr bulk crystals were grown using the chemical vapor transport method~\cite{klein_control_2022}. Thin CrSBr flakes utilized for measurements are derived from bulk crystals through the exfoliation method. This process entailed initially cleaving the crystals onto scotch tape to maintain a fixed crystal orientation. Subsequently, the crystals were mechanically exfoliated onto polydimethylsiloxane (PDMS). After this, a substrate SiO$_2$/Si chip facilitated the transfer of the CrSBr from the PDMS to the chip. The thickness of the CrSBr flakes was determined using atomic force microscopy (AFM).

\subsubsection*{Photoluminescence (PL) and reflectivity measurement}
Photoluminescence (PL) and reflectivity measurements were conducted on the CrSBr sample using the Montana cryostat at a base temperature of 4 K. 
For PL measurements, the sample was excited using a continuous wave (CW) diode laser at a wavelength of 532 nm. Reflectivity measurements utilized a broadband CW white light source. 
A 100X objective (Mitutoyo M Plan Apo) was employed to focus the incident light on the sample and to collect both the reflected light and the PL signal. This collected signal was subsequently analyzed using a spectrometer by Princeton Instruments (Model SpectraPro HRS-500) equipped with a grating of 300 lines per millimeter.

\subsubsection*{Magnetic field induced shifts}
Measurements of magnetic field-induced shifts were performed using a closed-cycle cryostat (attoDRY 2100) at a base temperature of 1.6 K. The magnetic field was oriented along the easy ($b$), intermediate ($a$), and hard ($c$) axes of the sample. Reflectivity measurements in the presence of a magnetic field employed a broadband continuous wave (CW) white light source. The input light passed through a linear polarizer to align its polarization with the $b$-axis.

\subsubsection*{Fluence dependence measurements}
Fluence dependence measurements were conducted using a closed-cycle cryostat (OptiCool) at a base temperature of $\sim$2~K (see Fig.~\ref{SIfig:diagram}). To mitigate the heating effects of continuous wave (CW) light sources, a pulsed white light laser, featuring a 50 ps pulse width and a 10 MHz repetition rate, was used to illuminate the sample for reflectivity measurements. The laser beam first passed through a tunable neutral density (ND) filter to adjust the power level. It was then filtered through a 690 nm long pass filter and a 740 nm short pass filter to narrow and specify the light spectrum. A linear polarizer aligned the laser's polarization with the b-axis. Finally, a 100X objective focused the laser onto the sample and collected the reflected signal.

\subsubsection*{Ultrafast transient reflection spectroscopy}
CrSBr flakes were mounted in a cryogenic chamber (Montana Instruments Cyrostation, s50) at a base temperature of 6K with a tunable magnetic field (Magneto-Optic module). 
Ultrafast transient reflection measurements were conducted using a Yb:KGW laser, operating at 90 kHz. A 600 nm pump with a $\sim$ 120 fs pulse width and broadband probe was focused onto the CrSBr surface using a 0.45 NA, 50X microscope objective (Nikon), resulting in 14 \textmu m pump and 2 \textmu m probe diameters (D4$\sigma$). 
Reflected probe pulses were collected with a fast line-scan camera (Teledyne e2V Octoplus USB), where a burst modulation scheme was employed in the shot-to-shot detection\cite{hall_optimizing_2023}.


\clearpage 


\section*{Acknowledgments}

\paragraph{Funding:}
B.D. and V.M.M. were supported by the Gordon and Betty Moore Foundation (Grant No. 12764), P.C.A. was supported by the Army Research Office grant W911NF-23-1-0394, S.Y. was supported by the NSF grant 2216838. A.V.D. was supported by the NSF grant DMR-2130544. A.V., F.K. and E.K. were supported by NSF grant OMA-2328993. S.A., D.P., and M.v.S were supported by the Computational Chemical Sciences program within the Office of Basic Energy Sciences, U.S. Department of Energy under Contract No. DE-AC36-08GO28308. 
S.A., D.P., and M.v.S acknowledge the use of the National Energy Research Scientific Computing
Center, under Contract No. DE-AC02-05CH11231 using NERSC award BES-ERCAP0021783 and the computational resources sponsored by the Department of Energy's Office of Energy
Efficiency and Renewable Energy and located at the National Renewable Energy Laboratory.
A.K. was supported by the Spanish Ministry for Science and Innovation– AEI grant CEX2023-001316-M (through the “Maria de Maeztu” Programme for Units of Excellence in R\&D) and grant RYC2021 031063-I, and by starting funds from the RPTU. A.B.K. was supported by the office of Naval Research grant N00014-24-1-2483. A.A., J.Q., and W.W. were supported by the Office of Naval Research and the Simons Foundation. S.J.H. acknowledges support from NSF Grant No. HRD-2112550 (Phase II CREST Center IDEALS). M.Y.S was supported by the Gordon and Betty Moore Foundation, grant DOI 10.37807/gbmf12235.

\paragraph{Author contributions:}
B.D., P.C.A., and V.M.M. conceived the experimental idea and interpreted the results together with all the authors. 
K.M. and Z.S. synthesized the CrSBr crystals. 
P.C.A., S.Y., and B.D. performed the experiments with help from A.V, F.K, E.K., J.Q., and W.W. supervised by A.B.K. and A.A. 
B.D. led the data analysis with help from P.C.A. and S.Y.
A.V.D and S.J.H. performed the pump-probe measurements and analyzed the corresponding data with help from M.Y.S.
S.A. performed the QSGW calculations in collaboration with D.P. and M.v.S. 
A.K. theoretically modeled the excitonic interactions. 
P.C.A., B.D., V.M.M., S.A., A.K., and S.Y. wrote the manuscript with input from all coauthors. 
V.M.M. supervised the project.

\paragraph{Competing interests:}
There are no competing interests to declare.

\paragraph{Data and materials availability:}
All data are available in the manuscript or the supplementary information.


\subsection*{Supplementary information}
Supplementary Text\\
Figs. S1 to S10\\
References \textit{(38-\arabic{enumiv})} 


\newpage


\renewcommand{\thefigure}{S\arabic{figure}}
\renewcommand{\thetable}{S\arabic{table}}
\renewcommand{\theequation}{S\arabic{equation}}
\renewcommand{\thepage}{S\arabic{page}}
\setcounter{figure}{0}
\setcounter{table}{0}
\setcounter{equation}{0}
\setcounter{page}{1} 
\setcounter{secnumdepth}{4}

\renewcommand{\theparagraph}{\Roman{paragraph}}


\begin{center}
\section*{Supplementary information for\\ \scititle}

Biswajit~Datta$^{\ast\dagger}$,
	Pratap~Chandra~Adak$^{\ast\dagger}$,
        Sichao~Yu$^{\dagger}$,\\
	Agneya~V.~Dharmapalan,
        Siedah~J.~Hall$^{}$,
        Anton~Vakulenko$^{}$,\\
        Filipp~Komissarenko$^{}$,
        Egor~Kurganov$^{}$, 
        Jiamin~Quan$^{}$,
        Wei~Wang$^{}$, \\
        Kseniia~Mosina$^{}$, 
        Zdeněk~Sofer$^{}$, 
        Dimitar~Pashov$^{}$, Mark~van~Schilfgaarde$^{}$,\\
        Swagata~Acharya$^{}$, 
        Akashdeep~Kamra$^{}$, 
        Matthew~Y.~Sfeir$^{}$,\\
        Andrea~Alù$^{}$, 
        Alexander~B.~Khanikaev$^{}$, 
        Vinod~M.~Menon$^{\ast}$\\ 
\small$^\ast$Corresponding authors. Email: bdatta@ccny.cuny.edu, padak@ccny.cuny.edu, vmenon@ccny.cuny.edu\\
\small$^\dagger$These authors contributed equally to this work.
\end{center}

\subsubsection*{This PDF file includes:}
Supplementary Text\\
Figures S1 to S10

\newpage


\subsection*{Supplementary Text}

\begin{figure}[h!]
    \centering
	\includegraphics[width=15.5cm]{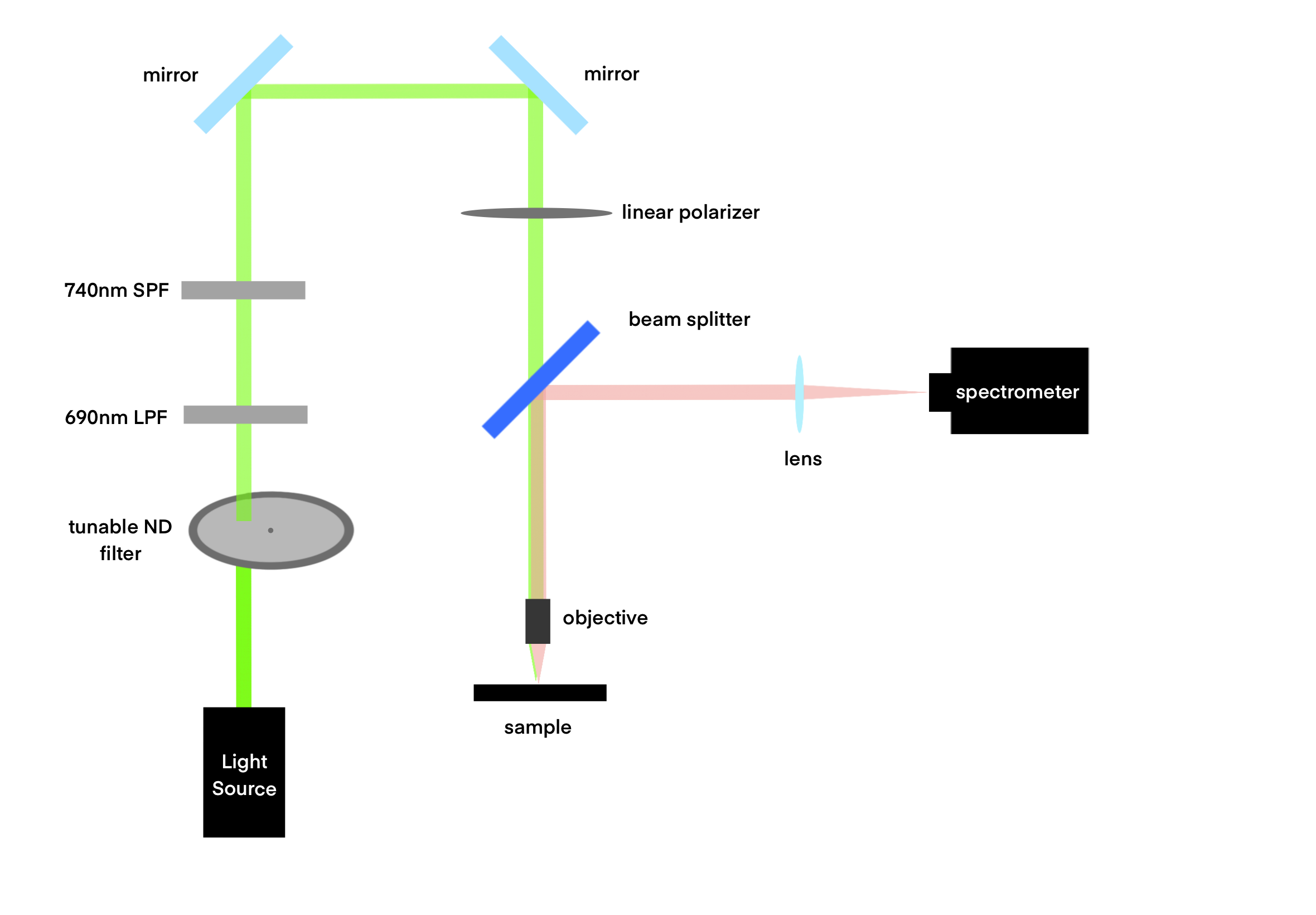}
        \caption{  \textbf{Schematic of the setup for power dependent measurements.}
        \textbf{}
        }
        \label{SIfig:diagram}
\end{figure}



\begin{figure}[hbt!]
    \centering
	\includegraphics[width=15.5cm]{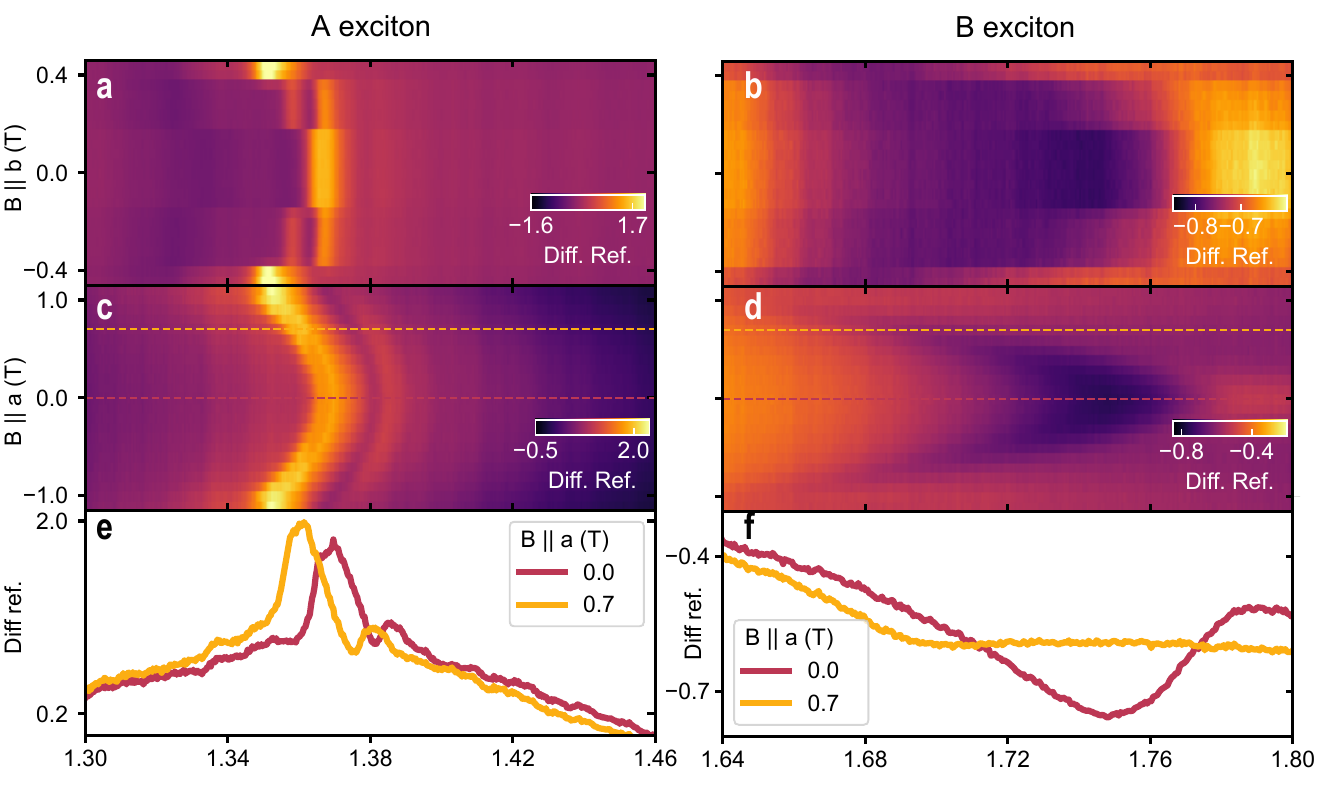}
        \caption{\textbf{In-plane magnetic field induced shifts in exciton energy.}
        \textbf{a,b},~Color-scale plots of the differential reflectivity as a function of energy and magnetic field applied along the $b$-axis for $A$ and $B$ excitons, respectively. In both cases, the exciton energy exhibits abrupt redshifts due to spin-flip transitions under the magnetic field.
        \textbf{c,d},~Similar color-scale plots for magnetic field applied along $a$-axis, showing a gradual energy shift due to progressive spin canting.
        \textbf{e,f},~Line plots of differential reflectivity vs. energy for two different magnetic field values applied along the $a$-axis.}
        \label{SIfig:ab_axis}
\end{figure}

\paragraph{Comparison of magnetic field dependence of $A$ and $B$ exciton.}

\begin{figure}[h!]
    \centering
	\includegraphics[width=15.5cm]{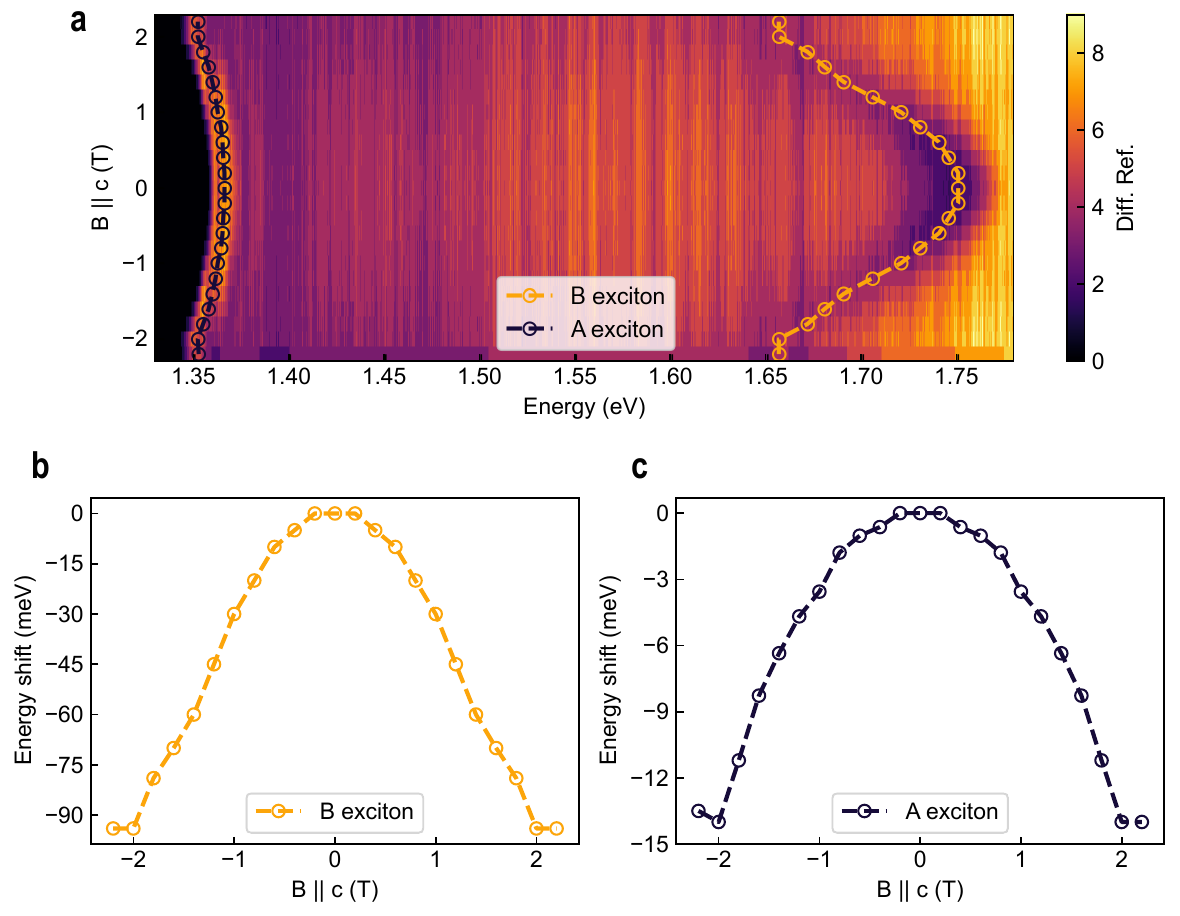}
        \caption{  \textbf{Out of plane magnetic field induced shifts in exciton energy.}
        \textbf{a},~Color-scale plots of the differential reflectivity as a function of energy and magnetic field applied along $c$-axis.
        \textbf{b,c},~Plots showing the magnetic field-induced energy shift of $B$ and $A$ excitons, respectively, extracted from \textbf{a}. 
        }
        \label{SIfig:caxis}
\end{figure}

Fig.~\ref{SIfig:caxis}a shows color-coded plots of differential reflectivity as a function of energy around $A$ exciton and $B$ exciton energy and magnetic field applied along the $c$-axis. Fig.~\ref{SIfig:caxis}b,c show how the exciton energy shifts of $A$ exciton and $B$ exciton change as a function of the magnetic field. We observed a continuous evolution of the exciton energy until reaching the saturation magnetic fields at $B_{c,\text{sat}}=2.0$~T for both $A$ exciton and $B$ exciton. This behavior is similar to the energy shift observed when the magnetic field is aligned along the $a$-axis, suggesting a consistent change in the canting angle from 
$\pi$ to 0 in both scenarios. $B$ exciton also shows a larger energy shift compared to $A$ exciton when the magnetic field is applied along the $c$-axis.

\paragraph{Additional fluence dependence data.}

\begin{figure}[h!]
    \centering
	\includegraphics[width=15cm]{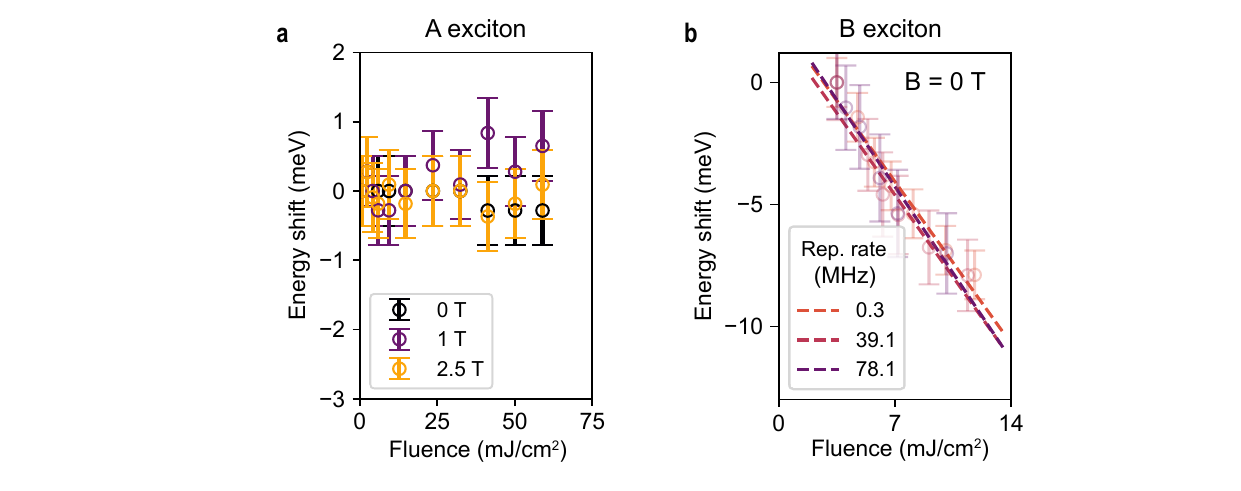}
        \caption{  \textbf{Fluence dependent measurement on $A$ exciton and repetition rate control for $B$ exciton.}
        \textbf{a},~Energy shift vs. fluence plots for the $A$ exciton, derived from resonant differential reflectivity measurements at different fields. The shift is minimal compared to that for $B$ exciton (main text).
        \textbf{b},~Energy shift vs. fluence plots for the $B$ exciton for three different laser repetition rates. The energy shift is independent of the repetition rate.
        }
        \label{SIfig:control}
\end{figure}

Our power-dependent measurements on the A exciton reveal a very small energy shift with varying laser pump power (Fig.~\ref{SIfig:control}a). This minor shift is likely due to the A exciton’s Frenkel character and its weaker exciton-magnon coupling compared to the B exciton. Additionally, measurements of the B exciton at different laser repetition rates show that the energy shift is independent of the repetition rate (Fig.~\ref{SIfig:control}b). Together, these observations rule out any  possibility of laser-induced temperature increase of the sample.

\begin{figure}[h]
    \centering
	\includegraphics[width=15.5cm]{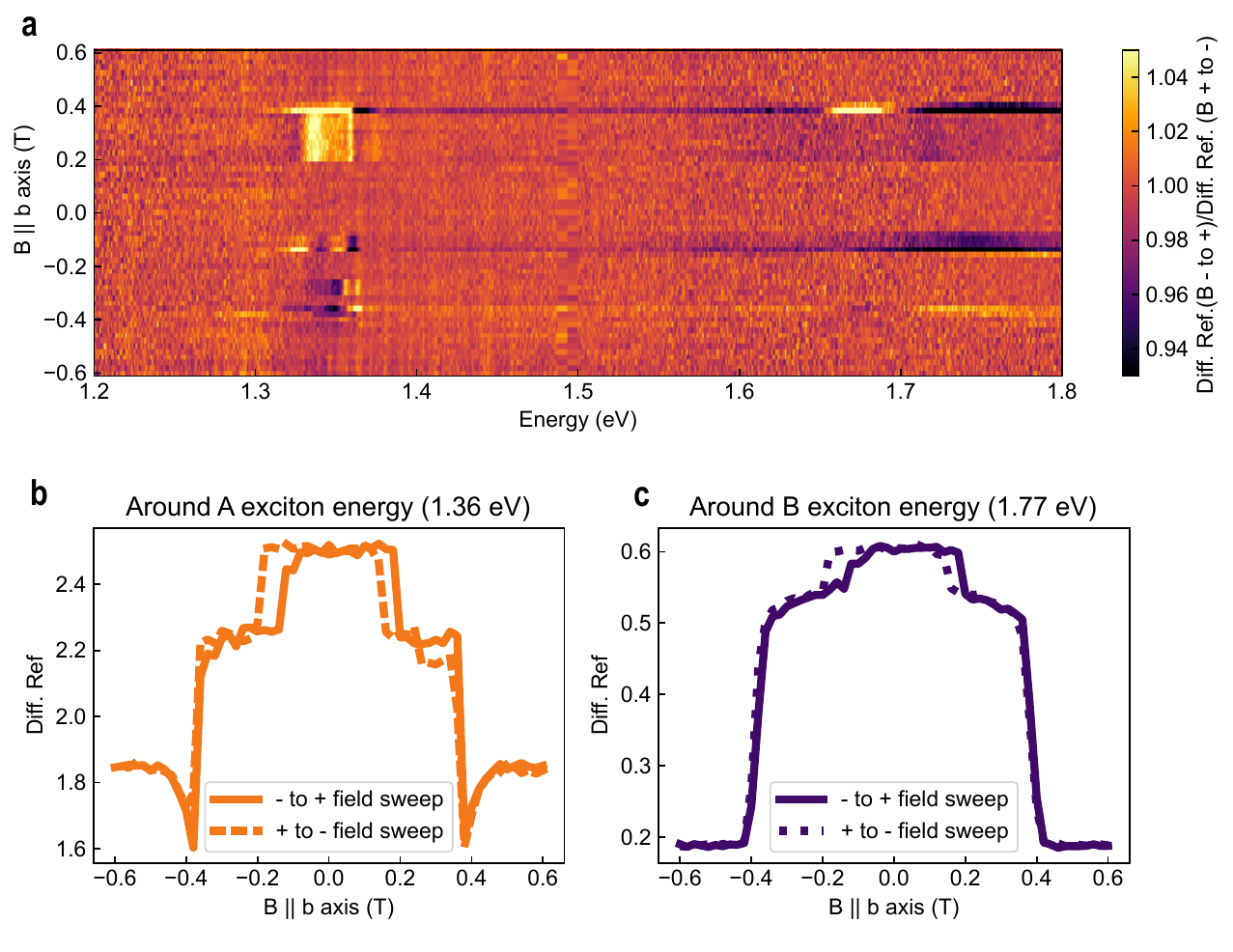}
        \caption{  \textbf{Hysteresis.}
        \textbf{a},~A color-scale plot of the ratios between differential reflectivity measured for sweeping the magnetic field in two directions. 
        \textbf{b,c},~Line plots showing the hysteresis in the differential reflectivity around the energy of $A$ and $B$ excitons, respectively.
        }
        \label{SIfig:hysteresis}
\end{figure}

\paragraph{Hysteresis.}
Forward and backward magnetic field sweeps show a pronounced hysteresis close to both A and B exciton resonances expected due to the underlying coupled magnetic order. Fig.~\ref{SIfig:hysteresis}a shows the ratio of the forward and backward differential reflection as a function of magnetic field and energy. Fig.~\ref{SIfig:hysteresis}b and Fig.~\ref{SIfig:hysteresis}c are the line cuts close to A and B exciton energy, respectively.

\clearpage

\paragraph{Details of theoretical calculation.}
The Quasiparticle Self-Consistent GW approximation~\cite{mark06qsgw,questaal_paper} is a self-consistent form of Hedin's GW approximation~\cite{hedin65}.  Self-consistency removes the starting point dependence, and as a result, the discrepancies are much more systematic than conventional forms of GW.  The great majority of such discrepancies in insulators originate from the omission of electron-hole interactions in the RPA polarizability.  By adding ladders to the RPA, electron-hole effects are taken into account.  Generating \textit{W} with ladder diagrams has important consequences;  screening is enhanced and \textit{W} reduced.  This in turn reduces fundamental bandgaps and also valence bandwidths.  Agreement with experiments in both one-particle and two-particle properties is greatly improved. The theory and its application to a large number of both weakly and strongly correlated insulators is given in Ref.~\cite{Cunningham2023}. The importance of self-consistency in both $\mathrm{QSGW}$ and $\mathrm{QSG\hat{W}}$ for different materials have been explored \cite{acharya2021importance}.

For bulk CrSBr in the AFM phase with a 12-atom unit cell, we use a=3.504 \AA, b=4.738 \AA. Individual layers contain ferromagnetically polarized spins pointing either along the $+b$ or $-b$ axis, while the interlayer coupling is antiferromagnetic. The single particle calculations (LDA and energy band calculations with the static quasiparticlized $\mathrm{QSGW}$ and $\mathrm{QSG\hat{W}}$ $\Sigma$(k)) are performed on a 10$\times$7$\times$2 k-mesh while the relatively smooth dynamical self-energy $\Sigma(\omega)$ is constructed using a 6$\times$4$\times$2 k-mesh.  The $\mathrm{QSGW}$ and $\mathrm{QSG\hat{W}}$ cycles are iterated until the RMS change in the static part of quasiparticlized self-energy $\Sigma(0)$  reaches 10$^{-5}$ Ry. The two-particle Hamiltonian that is solved self-consistently to compute both the $\Sigma$ and the excitonic eigenvalues and eigenfunctions contained 26 valence bands and 9 conduction bands. Excitonic eigenvalues of the two-particle Hamiltonian are converged using 10$\times$7$\times$2 k-mesh.

\begin{figure}[h!]
    \centering
	\includegraphics[width=15.5cm]{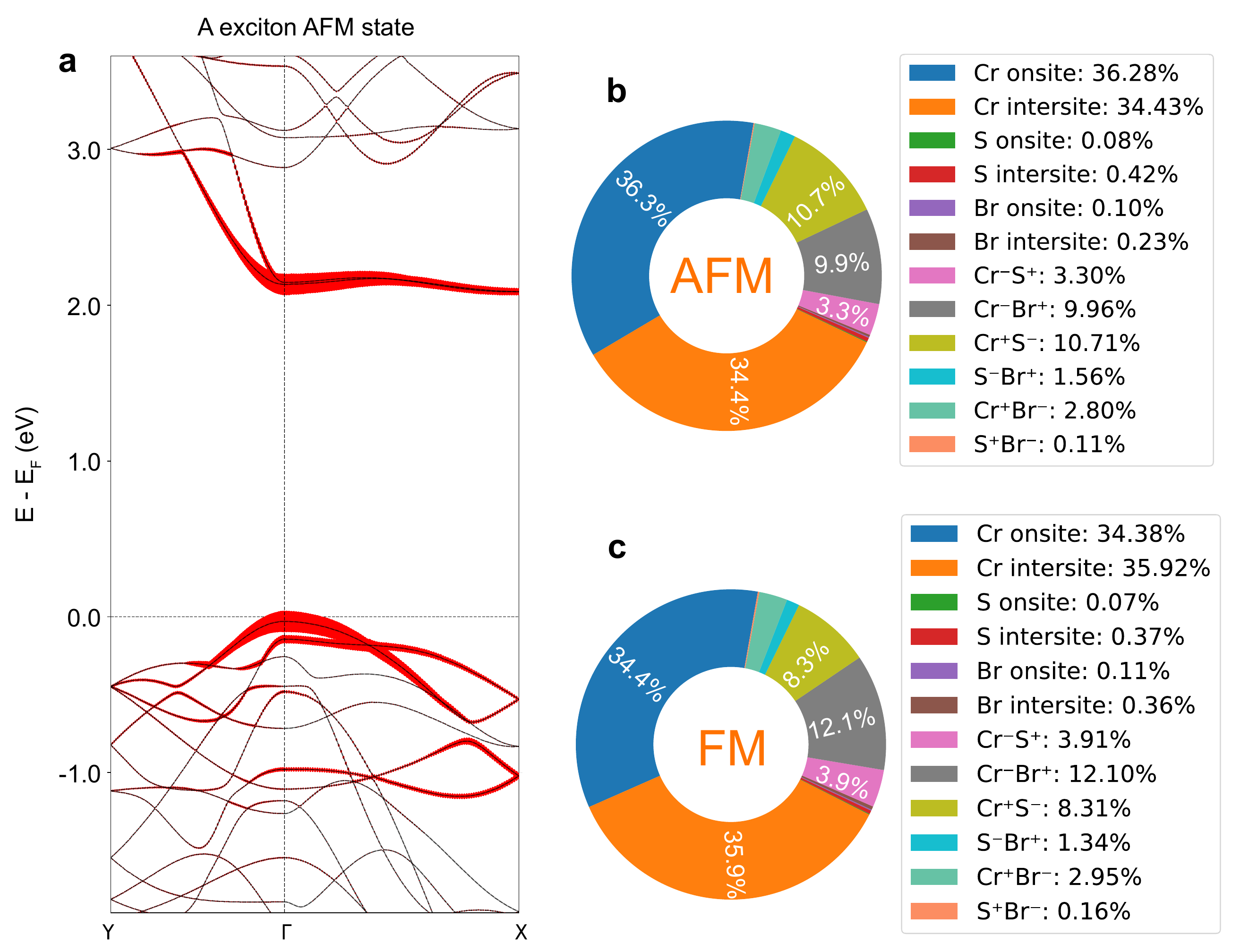}
        \caption{  \textbf{$A$ exciton.} \textbf{a},~Highlighted momentum states of the band structure responsible for the $A$ exciton in the antiferromagnetic (AFM) phase.
        \textbf{b},~The different orbitals involved in forming the $A$ exciton in the AFM phase.
        \textbf{c},~The different orbitals involved in forming the $A$ exciton in the ferromagnetic (FM) phase.
        }
        \label{SIfig:bandstructure_A}
\end{figure}

\begin{figure}[h!]
    \centering
	\includegraphics[width=15.5cm]{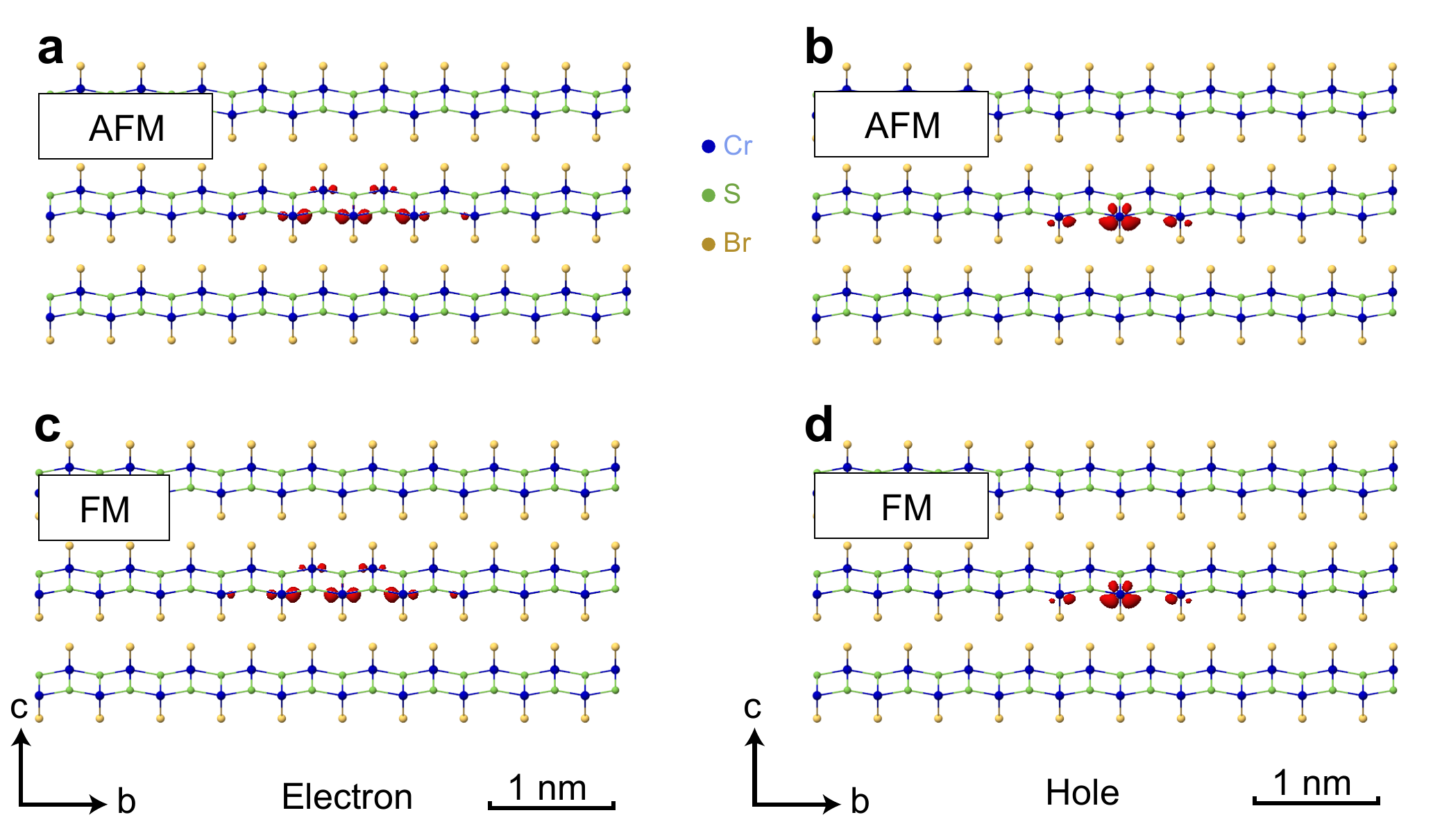}
        \caption{  \textbf{Wavefunction of the $A$ exciton.} \textbf{a} and \textbf{c}, Electron wavefunction
        in AFM and FM phases, respectively. 
        \textbf{b} and \textbf{d}, Hole wavefunction
        in AFM and FM phases, respectively. 
        }
        \label{SIfig:wavefunction_A}
\end{figure}

\begin{figure}[h!]
    \centering
	\includegraphics[width=15.5cm]{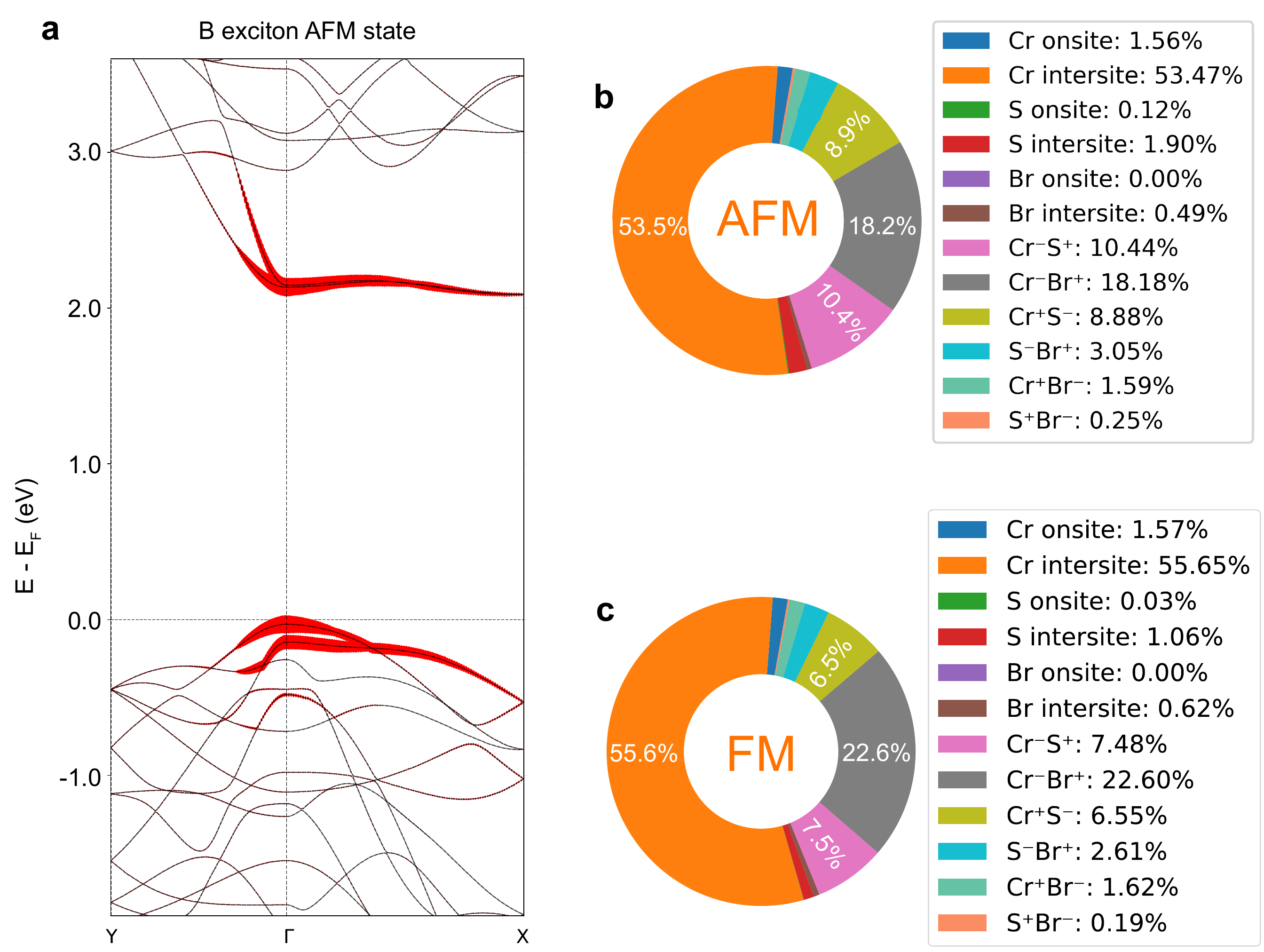}
        \caption{  \textbf{$B$ exciton.}
        \textbf{a},~Highlighted momentum states of the band structure responsible for the $B$ exciton in the antiferromagnetic (AFM) phase.
        \textbf{b},~The different orbitals involved in forming the $B$ exciton in the AFM phase.
        \textbf{c},~The different orbitals involved in forming the $B$ exciton in the ferromagnetic (FM) phase.}
\label{SIfig:bandstructure_B}
\end{figure}

\begin{figure}[h!]
    \centering
	\includegraphics[width=15.5cm]{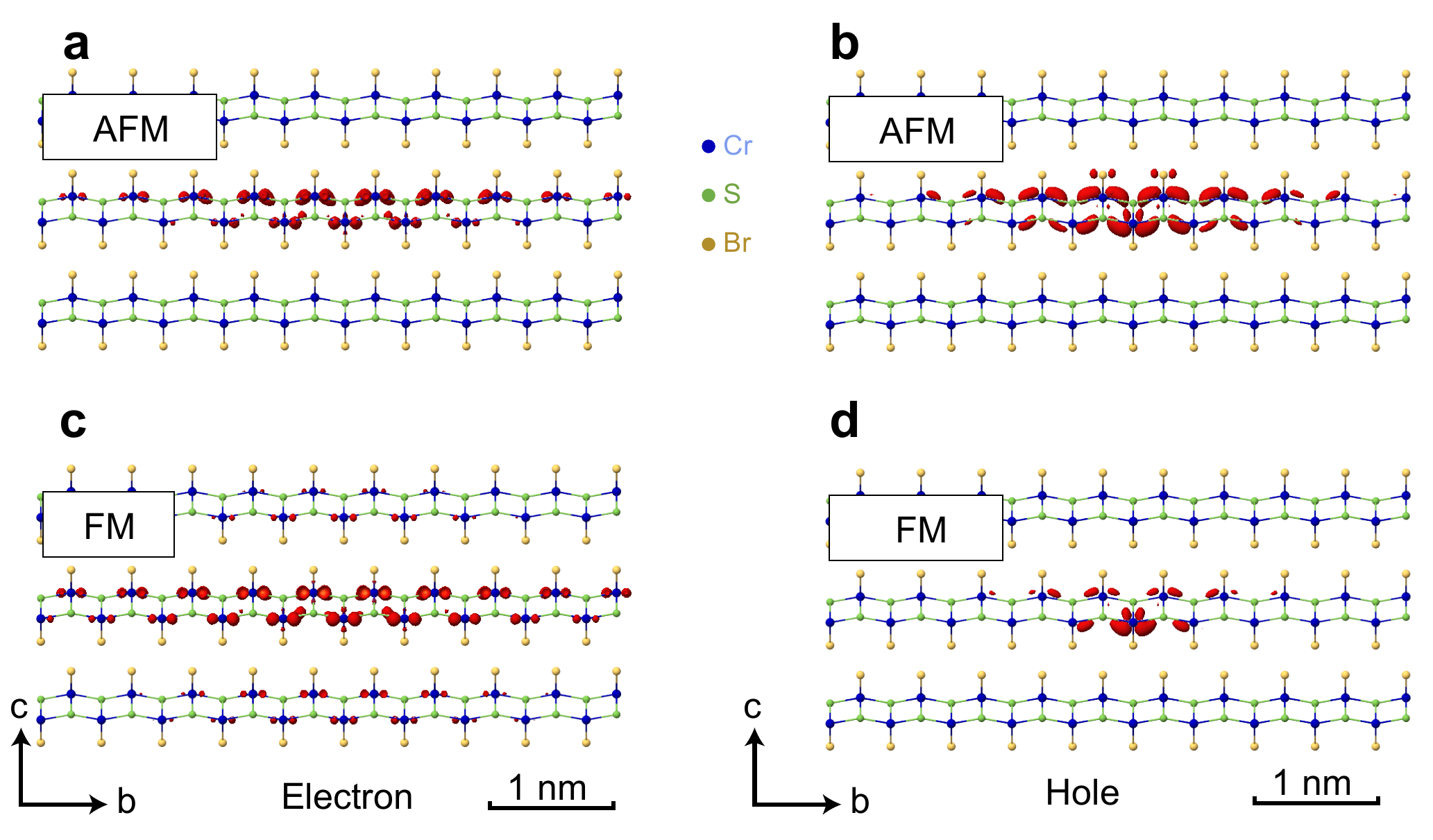}
        \caption{  \textbf{Wavefunction of the $B$ exciton.} \textbf{a} and \textbf{c}, Electron wavefunction
        in AFM and FM phases, respectively. 
        \textbf{b} and \textbf{d}, Hole wavefunction
        in AFM and FM phases, respectively. 
        }
        \label{SIfig:wavefunction_B}
\end{figure}

\clearpage

\paragraph{Theoretical model of magnon-mediated excitonic interactions.}

\begin{figure}[h!]
    \centering
	\includegraphics[width=15.5cm]{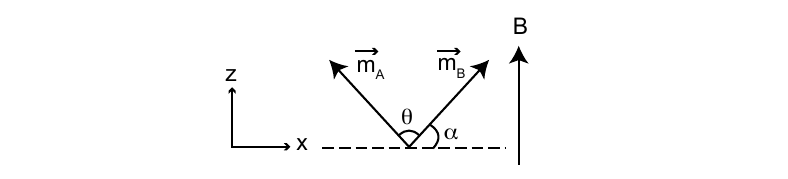}
        \caption{Schematic depiction of the two sublattice magnetizations in equilibrium. $\theta = 0$ corresponds to the ferromagnetic configuration, while $\theta = \pi$ corresponds to the antiferromagnetic configuration.
        }
        \label{SIfig:magnon_model}
\end{figure}

Here, we examine a particular contribution to magnon-mediated attractive exciton-exciton interaction that should be operational in CrSBr. 
The key physics and phenomenology are similar to the recent report of such an attraction mediated by phonons~\cite{yazdaniCouplingOctahedralTilts2024}. We also highlight a potentially important difference in terms of the different response times of lattice and magnetic orders. 
Our considerations indicate that the attractive exciton-exciton interaction is nonzero in the canted magnetic state of CrSBr, i.e., when the two sublattice magnetizations subtend an angle different from 0 and $\pi$ with each other in equilibrium.
In analogy with Ref.~\cite{yazdaniCouplingOctahedralTilts2024}, let us write down the energy of the magnetic plus excitonic system in CrSBr at a given applied magnetic field $B$:

\begin{align}
    E_{\text{total}}(N) &= E_{\text{cons}}(B) + \frac{1}{2}\chi\beta^2 + N E_{\text{X}}, \tag{1} \\
    &= E_{\text{cons}}(B) + \frac{1}{2}\chi\beta^2 + N \left( E_{\text{X},0} - \Delta_B \cos^2\left(\frac{\theta + \beta}{2}\right) \right), \tag{2}
\end{align}
where $N$ and $E_X$ are the number and energy of excitons, and $E_\text{cons}(B)$ captures all contributions to the energy for a given applied magnetic field that we do not explicitly consider here.
The angle $\beta$ characterizes the change in canting between the two sublattice magnetizations due to the spatially uniform magnon mode considered here. 
In this sense, $\beta$ can be seen as the small change in $\theta$, and we assume $|\beta| \ll |\theta|$. 
$\chi$ characterizes an energy corresponding to the considered magnon mode, the energy contribution of which is expressed via the $\beta^2$ term, $\Delta_B$ parametrizes the experimentally observed dependence of the exciton energy on the magnetization canting in CrSBr~\cite{wilson_interlayer_2021, dirnberger_magneto-optics_2023}. 
We have considered only one magnon mode here to capture the key physics, in analogy with Ref.~\cite{yazdaniCouplingOctahedralTilts2024}. A more detailed theoretical model accounting for all the relevant magnon modes is expected to yield similar results and is left for future work.
Employing Eq.~(2) and the condition $|\beta| \ll |\theta|$, we may simplify the energy to
\begin{equation}
E_{\text{total}}(N) = E_{\text{cons}}(B) + N E_{\text{X},0} - N \Delta_B \cos^2\left(\frac{\theta}{2}\right) + \frac{1}{2} \chi \beta^2 + \frac{N \Delta_B}{2} \sin(\theta) \beta. \tag{3}
\end{equation}

Now, let us assume that the magnetization dynamics characterized by $\beta$ respond sufficiently fast to the number of excitons in the system. This assumption has been employed and seems to be experimentally validated for the case of exciton-phonon interaction~\cite{yazdaniCouplingOctahedralTilts2024}. 
Whether this assumption, which seems reasonable for high-energy optical phonons, is valid for magnetization dynamics or not remains unclear at present and can be adequately examined in a detailed model accounting for all the relevant magnon modes. 
For now, within this assumption, the magnetic configuration characterized by $\beta$ adapts to the number of excitons in the system and minimizes its energy:
\begin{equation}
\frac{\partial}{\partial \beta} E_{\text{total}}(N) = 0, \tag{4}
\end{equation}

\begin{equation}
\implies \beta_{\text{min}} = -\frac{N \Delta_B}{2 \chi} \sin(\theta), \tag{5}
\end{equation}

\begin{equation}
\text{and} \quad E_{\text{min}}(N) = E_{\text{cons}}(B) + N E_{\text{X},0} - N \Delta_B \cos^2\left(\frac{\theta}{2}\right) - \frac{N^2 \Delta_B^2 \sin^2 \theta}{8 \chi}, \tag{6}
\end{equation}
where $\beta_{\text{min}}$ is the value of $\beta$, which minimizes the energy for a given number N of excitons, and $E_{\text{min}}$ is the corresponding minimum energy.
Equation (6) becomes our main result that captures the magnon-mediated exciton-exciton attraction, represented by the term \(\propto N^2\). 
To see this clearly, we evaluate the emitted photon energy when the system goes from a state with \(N\) to \(N - 1\) excitons:

\begin{align}
\hbar \omega &= E_{\min}(N) - E_{\min}(N-1), \tag{7} \\
&= E_{\text{X},0} - \Delta_B \cos^2 \left( \frac{\theta}{2} \right) - (2N - 1) \frac{\Delta_B^2 \sin^2 \theta}{8\chi}. \tag{8}
\end{align}

Here, the last term represents an exciton number-dependent red shift in the emission. 
We further note that this contribution vanishes for $\theta = 0$ or $\theta = \pi$, and thus requires a finite canting to be operational.




\begin{thebibliography}{10}
\providecommand{\url}[1]{\texttt{#1}}
\expandafter\ifx\csname urlstyle\endcsname\relax
  \providecommand{\doi}[1]{doi:\discretionary{}{}{}#1}\else
  \providecommand{\doi}{doi:\discretionary{}{}{}\begingroup \urlstyle{rm}\Url}\fi

\bibitem{reganEmergingExcitonPhysics2022b}
E.~C. Regan, \emph{et~al.}, Emerging Exciton Physics in Transition Metal Dichalcogenide Heterobilayers. \emph{Nature Reviews Materials} \textbf{7}~(10), 778--795 (2022).

\bibitem{snokeSpontaneousBoseCoherence2002}
D.~Snoke, Spontaneous {{Bose Coherence}} of {{Excitons}} and {{Polaritons}}. \emph{Science} \textbf{298}~(5597), 1368--1372 (2002).

\bibitem{wangEvidenceHightemperatureExciton2019b}
Z.~Wang, \emph{et~al.}, Evidence of High-Temperature Exciton Condensation in Two-Dimensional Atomic Double Layers. \emph{Nature} \textbf{574}~(7776), 76--80 (2019).

\bibitem{dengCondensationSemiconductorMicrocavity2002}
H.~Deng, G.~Weihs, C.~Santori, J.~Bloch, Y.~Yamamoto, Condensation of {{Semiconductor Microcavity Exciton Polaritons}}. \emph{Science} \textbf{298}~(5591), 199--202 (2002).

\bibitem{kasprzakBoseEinsteinCondensation2006a}
J.~Kasprzak, \emph{et~al.}, Bose--{{Einstein}} Condensation of Exciton Polaritons. \emph{Nature} \textbf{443}~(7110), 409--414 (2006).

\bibitem{amoSuperfluidityPolaritonsSemiconductor2009}
A.~Amo, \emph{et~al.}, Superfluidity of Polaritons in Semiconductor Microcavities. \emph{Nature Physics} \textbf{5}~(11), 805--810 (2009).

\bibitem{liExcitonicSuperfluidPhase2017a}
J.~I.~A. Li, T.~Taniguchi, K.~Watanabe, J.~Hone, C.~R. Dean, Excitonic Superfluid Phase in Double Bilayer Graphene. \emph{Nature Physics} \textbf{13}~(8), 751--755 (2017).

\bibitem{jeromeExcitonicInsulator1967}
D.~J{\'e}rome, T.~M. Rice, W.~Kohn, Excitonic {{Insulator}}. \emph{Physical Review} \textbf{158}~(2), 462--475 (1967).

\bibitem{cercellierEvidenceExcitonicInsulator2007}
H.~Cercellier, \emph{et~al.}, Evidence for an {{Excitonic Insulator Phase}} in ${{1T}}-\text{TiSe}_2$. \emph{Physical Review Letters} \textbf{99}~(14), 146403 (2007).

\bibitem{kogarSignaturesExcitonCondensation2017}
A.~Kogar, \emph{et~al.}, Signatures of Exciton Condensation in a Transition Metal Dichalcogenide. \emph{Science} \textbf{358}~(6368), 1314--1317 (2017).

\bibitem{joglekarWignerSupersolidExcitons2006}
Y.~N. Joglekar, A.~V. Balatsky, S.~Das~Sarma, Wigner Supersolid of Excitons in Electron-Hole Bilayers. \emph{Physical Review B} \textbf{74}~(23), 233302 (2006).

\bibitem{johansen_magnon-mediated_2019}
O.~Johansen, A.~Kamra, C.~Ulloa, A.~Brataas, R.~A. Duine, Magnon-Mediated Indirect Exciton Condensation through Antiferromagnetic Insulators. \emph{Physical Review Letters} \textbf{123}~(16), 167203 (2019).

\bibitem{axtNonlinearOpticsSemiconductor1998}
V.~M. Axt, S.~Mukamel, Nonlinear Optics of Semiconductor and Molecular Nanostructures; a Common Perspective. \emph{Reviews of Modern Physics} \textbf{70}~(1), 145--174 (1998).

\bibitem{liDipolarInteractionsLocalized2020}
W.~Li, X.~Lu, S.~Dubey, L.~Devenica, A.~Srivastava, Dipolar Interactions between Localized Interlayer Excitons in van Der {{Waals}} Heterostructures. \emph{Nature Materials} \textbf{19}~(6), 624--629 (2020).

\bibitem{wilson_interlayer_2021}
N.~P. Wilson, \emph{et~al.}, Interlayer electronic coupling on demand in a {2D} magnetic semiconductor. \emph{Nature Materials} \textbf{20}~(12), 1657--1662 (2021).

\bibitem{bae_exciton-coupled_2022}
Y.~J. Bae, \emph{et~al.}, Exciton-coupled coherent magnons in a {2D} semiconductor. \emph{Nature} \textbf{609}~(7926), 282--286 (2022).

\bibitem{diederich_tunable_2023}
G.~M. Diederich, \emph{et~al.}, Tunable interaction between excitons and hybridized magnons in a layered semiconductor. \emph{Nature Nanotechnology} \textbf{18}~(1), 23--28 (2023).

\bibitem{dirnberger_magneto-optics_2023}
F.~Dirnberger, \emph{et~al.}, Magneto-optics in a van der {Waals} magnet tuned by self-hybridized polaritons. \emph{Nature} \textbf{620}~(7974), 533--537 (2023).

\bibitem{brennan_important_2024}
N.~J. Brennan, C.~A. Noble, J.~Tang, M.~E. Ziebel, Y.~J. Bae, Important {Elements} of {Spin}-{Exciton} and {Magnon}-{Exciton} {Coupling}. \emph{ACS Physical Chemistry Au} \textbf{4}~(4), 322--327 (2024).

\bibitem{telford_layered_2020}
E.~J. Telford, \emph{et~al.}, Layered {Antiferromagnetism} {Induces} {Large} {Negative} {Magnetoresistance} in the van der {Waals} {Semiconductor} {CrSBr}. \emph{Advanced Materials} \textbf{32}~(37), 2003240 (2020).

\bibitem{wang_electrically-tunable_2020}
H.~Wang, J.~Qi, X.~Qian, Electrically-{Tunable} {High} {Curie} {Temperature} {Two}-{Dimensional} {Ferromagnetism} in {Van} der {Waals} {Layered} {Crystals}. \emph{Applied Physics Letters} \textbf{117}~(8), 083102 (2020).

\bibitem{lee_magnetic_2021}
K.~Lee, \emph{et~al.}, Magnetic Order and Symmetry in the {2D} Semiconductor {CrSBr}. \emph{Nano Letters} \textbf{21}~(8), 3511--3517 (2021).

\bibitem{scheie_spin_2022}
A.~Scheie, \emph{et~al.}, Spin {Waves} and {Magnetic} {Exchange} {Hamiltonian} in {CrSBr}. \emph{Advanced Science} \textbf{9}~(25), 2202467 (2022).

\bibitem{PhysRevB.107.235107}
M.~Bianchi, \emph{et~al.}, Paramagnetic electronic structure of {CrSBr}: Comparison between $ab$ $initio$ ${GW}$ theory and angle-resolved photoemission spectroscopy. \emph{Phys. Rev. B} \textbf{107}, 235107 (2023).

\bibitem{watson2024giant}
M.~D. Watson, \emph{et~al.}, Giant exchange splitting in the electronic structure of {A}-type {2D} antiferromagnet {CrSBr}. \emph{npj 2D Materials and Applications} \textbf{8}~(1), 1--8 (2024).

\bibitem{shao2024exciton}
Y.~Shao, \emph{et~al.}, Magnetically confined surface and bulk excitons in a layered antiferromagnet. \emph{Under review}  (2024).

\bibitem{wang_magnetically-dressed_2023}
T.~Wang, \emph{et~al.}, Magnetically-dressed {CrSBr} exciton-polaritons in ultrastrong coupling regime. \emph{Nature Communications} \textbf{14}~(1), 5966 (2023).

\bibitem{sun_dipolar_2024}
Y.~Sun, \emph{et~al.}, Dipolar spin wave packet transport in a van der {Waals} antiferromagnet. \emph{Nature Physics} \textbf{20}~(5), 794--800 (2024).

\bibitem{guo_chromium_2018}
Y.~Guo, Y.~Zhang, S.~Yuan, B.~Wang, J.~Wang, Chromium sulfide halide monolayers: intrinsic ferromagnetic semiconductors with large spin polarization and high carrier mobility. \emph{Nanoscale} \textbf{10}~(37), 18036--18042 (2018).

\bibitem{methods}
Materials and methods are available as supplementary material.

\bibitem{datta2022highly}
B.~Datta, \emph{et~al.}, Highly nonlinear dipolar exciton-polaritons in bilayer MoS2. \emph{Nature communications} \textbf{13}~(1), 6341 (2022).

\bibitem{louca2023interspecies}
C.~Louca, \emph{et~al.}, Interspecies exciton interactions lead to enhanced nonlinearity of dipolar excitons and polaritons in {MoS$_2$} homobilayers. \emph{Nature Communications} \textbf{14}~(1), 3818 (2023).

\bibitem{Cunningham2023}
B.~Cunningham, M.~Gr{\"u}ning, D.~Pashov, M.~van Schilfgaarde, {QS$G\hat{W}$: Quasiparticle Self Consistent $GW$ with Ladder Diagrams in $W$}. \emph{Phys. Rev. B} \textbf{108}, 165104 (2023).

\bibitem{yazdaniCouplingOctahedralTilts2024}
N.~Yazdani, \emph{et~al.}, Coupling to Octahedral Tilts in Halide Perovskite Nanocrystals Induces Phonon-Mediated Attractive Interactions between Excitons. \emph{Nature Physics} \textbf{20}~(1), 47--53 (2024).

\bibitem{chaves_tunable_2021}
A.~Chaves, F.~M. Peeters, Tunable effective masses of magneto-excitons in two-dimensional materials. \emph{Solid State Communications} \textbf{334-335}, 114371 (2021).

\bibitem{zipfel_spatial_2018}
J.~Zipfel, \emph{et~al.}, Spatial extent of the excited exciton states in {WS}$_2$ monolayers from diamagnetic shifts. \emph{Physical Review B} \textbf{98}~(7), 075438 (2018).

\bibitem{pacuski_excitonic_2006}
W.~Pacuski, \emph{et~al.}, Excitonic {Giant} {Zeeman} {Effect} in {Wide} {Gap} {Diluted} {Magnetic} {Semiconductors} {Based} on {ZnO} and {GaN}. \emph{Acta Physica Polonica A} \textbf{110}~(3), 303--309 (2006).

\bibitem{klein_control_2022}
J.~Klein, \emph{et~al.}, Control of structure and spin texture in the van der {Waals} layered magnet {CrSBr}. \emph{Nature Communications} \textbf{13}~(1), 5420 (2022).

\bibitem{hall_optimizing_2023}
S.~J. Hall, P.~J. Budden, A.~Zats, M.~Y. Sfeir, Optimizing the sensitivity of high repetition rate broadband transient optical spectroscopy with modified shot-to-shot detection. \emph{Review of Scientific Instruments} \textbf{94}~(4), 043005 (2023).

\bibitem{mark06qsgw}
M.~van Schilfgaarde, T.~Kotani, S.~Faleev, {Quasiparticle Self-Consistent \emph{GW} Theory}. \emph{Phys. Rev. Lett.} \textbf{96}~(22), 226402 (2006).

\bibitem{questaal_paper}
D.~Pashov, \emph{et~al.}, {Questaal: a package of electronic structure methods based on the linear muffin-tin orbital technique}. \emph{Comp. Phys. Comm.} \textbf{249}, 107065 (2020).

\bibitem{hedin65}
L.~Hedin, {New Method for Calculating the One-Particle Green's Function with Application to the Electron-Gas Problem}. \emph{Phys. Rev.} \textbf{139}, A796 (1965).

\bibitem{acharya2021importance}
S.~Acharya, \emph{et~al.}, Importance of charge self-consistency in first-principles description of strongly correlated systems. \emph{npj Computational Materials} \textbf{7}~(1), 208 (2021).

\end{thebibliography}
\end{document}